\documentclass[conference]{IEEEtran}
\IEEEoverridecommandlockouts
\usepackage{amsmath}
\usepackage{amsfonts}
\usepackage{graphicx}
\usepackage{epstopdf}
\usepackage{xcolor}
\usepackage{cite}
\usepackage{comment}
\usepackage{cases}
\def\BibTeX{{\rm B\kern-.05em{\sc i\kern-.025em b}\kern-.08em
    T\kern-.1667em\lower.7ex\hbox{E}\kern-.125emX}}

\begin{document}

\title{ {\small Accepted in 2024 IEEE PES GENERAL MEETING, Seattle, Washington (PES GM 2024)}\\  
\textbf{Distributed  {Anomaly} Detection in Modern Power Systems: A Penalty-based Mitigation Approach } \thanks{This work is supported in part by National Science Foundation under Grants DMS-2229109.}}

\author{\IEEEauthorblockN{Erfan Mehdipour Abadi, Masoud H. Nazari~\IEEEmembership{Senior Member,~IEEE,}}
\\\small Department of Electrical and Computer Engineering, Wayne State University, Detroit, MI, USA
\\\small  erfan.abadi@wayne.edu, masoud.nazari@wayne.edu}

\maketitle

\begin{abstract}
The evolving landscape of electric power networks, influenced by the integration of 
 {distributed energy resources} require the development of novel power system monitoring and control architectures.
This paper develops algorithm to monitor and detect  {anomalies} of different parts of a power system that cannot be measured directly, by applying neighboring measurements and a dynamic probing technique in a distributed fashion.  {Additionally, the proposed method accurately assesses the severity of the anomaly. }
A decision-making algorithm is introduced to effectively penalize anomalous agents, ensuring vigilant oversight of the entire power system's functioning.
Simulation results show the efficacy of algorithms in distributed anomaly detection and mitigation.  
\end{abstract}

\section{Introduction} \label{Section1}

Climate change, rising fossil fuel costs, technological advances, and the growth of distributed energy resources (DER) make it necessary to re-evaluate the power systems control, monitoring, and security \cite{basak2012literature}. 
The goal is to ensure stability, optimize operations, and maintain security in this changing landscape. This requires innovative strategies to integrate and utilize distributed resources while upholding the reliability and resilience of modern power systems (MPS).

Traditional power networks follow a central structure, with control centers and supervisory control and data acquisition (SCADA) systems responsible for overseeing electricity production and distribution \cite{wu2006overview}. 
However, this structure suffers from drawbacks like vulnerability to single points of failure and scalability issues with integrating numerous DERs. Consequently, distributed monitoring and decision-making systems are growing in MPS \cite{georgilakis2015review}, leading a shift from centralized to decentralized management and control \cite{ackermann2001distributed}.



Transitioning to decentralization involves viewing the network as homogeneous, with agents known as prosumers who both produce and consume electricity. This structure ensures no single agent has direct control over others \cite{grijalva2011prosumer}. Despite benefits like reducing single-point-of-failure risks and enabling stable operation, the absence of central control may expose networks to anomalies. Thus, distributed monitoring of prosumers and detecting anomalies pose critical challenges.

Some anomalies in MPS could be due to the presence of malicious agents.  {Anomalies cover a broad range of abnormal operations. 
The focus of this paper is malicious anomalies which could either be caused by cyber-attacks such as false data injection \cite{EMS1} or intentional anomalies of the prosumers. Despite their distinction, they are related concepts that both cause prosumers to have a deviation from their expected optimal power values.} Malicious agents may object to disrupting power systems by operating them inefficiently \cite{li2018robust}, degrading devices, or destabilizing the system \cite{stamp2009reliability}. For instance, information modification in the power system of Puerto Rico has caused a loss of \$400 million in a year \cite{krebs2012fbi}.

Conversely, some prosumers may manipulate shared variable values to boost their financial gains \cite{alkhraijah2022analyzing} or accidentally; with this behavior persisting for an extended period without returning to the expected operational state. Addressing such anomalies requires imposing penalties on the involved prosumers. Accordingly, we classify anomalies into two categories: 1) malicious anomaly and, 2) accidental anomaly.

Through the application of the distributed optimal power flow (DOPF) technique, a stable and optimal operating point is established. Due to the lack of central monitoring, the presence of an anomalous agent may lead to disruptive consequences, impeding the normal operation of the power grid \cite{kim2000comparison, alkhraijah2022analyzing, du2022decentralized}. 

If the distributed network actions cannot be effectively policed and the anomaly is not detected and mitigated, stable and optimal operation of MPS cannot be guaranteed. Detecting anomaly in a distributed fashion has been explored in different large-scale dynamic systems such as vehicular ad hoc networks, wireless sensor networks, and mobile networks \cite{ghosh2009distributed,toledo2007robust,mahmood2023towards}. However, to the best of our knowledge, little effort has been made toward distributed anomaly detection in MPS. In \cite{EMS1}, a distributed energy management algorithm has been proposed for resilience under fault attacks and random attacks. 
\cite{Consensus1} proposes the detection of the adversary based on weight update value changes and isolation of the anomalous agents from economic dispatch schedules. 

Understanding anomalies in the power system goes beyond identification; grasping their severity, duration, and pattern is crucial for determining appropriate responses. 
This paper models anomalies as perturbations on the system's equilibrium, causing mismatches between agreed-upon and actual power production by prosumers, detected through dynamic probing with available infrastructures like line power flow measurements. It proposes distributed detection using dynamic probing to detect anomalous prosumers and a mitigation method is introduced based on the "anomalous factor" and penalty value functions.
%
%
The contributions of the paper are as follows:
\begin{itemize}
    \item Implementing a distributed monitoring scheme using a neighbor-watch-based approach, where neighbors utilize dynamic probing techniques for calculation of the  {energy mismatch} of the anomalous adjacent prosumers.
    \item Introducing an automated system for detecting and mitigating anomalies based on their  {energy mismatch}, aimed at calculating appropriate penalty values for errant prosumers and isolating those who persist in their misconduct.
\end{itemize}

The rest of the paper is organized as follows. In Section \ref{Section2} a distributed monitoring scheme based on dynamic probing is proposed. In section \ref{Section3}, an automated method for anomaly mitigation is introduced. Section \ref{Section4} illustrates the effectiveness of the proposed algorithm via simulation. The paper is concluded in Section \ref{Section5} with discussions on overall findings.

\section{Distributed Prosumer Monitoring} \label{Section2}
Prosumer in the power network is an abstraction of several devices such as generators, loads, storage devices, and lines. To apply a decentralized control scheme to the prosumer-based MPS, we must first create a model to embody the behavior of prosumers.
Numerous mathematical decomposition algorithms for non-convex nonlinear programming (NLP) problems have been applied to the OPF problem for individual nodes to achieve a DOPF \cite{ kim2000comparison, xie2021impact, xie2023learning}.

\begin{figure}[t]
\centering
\includegraphics[width=1\linewidth, trim={0 0cm 0 0cm},clip]{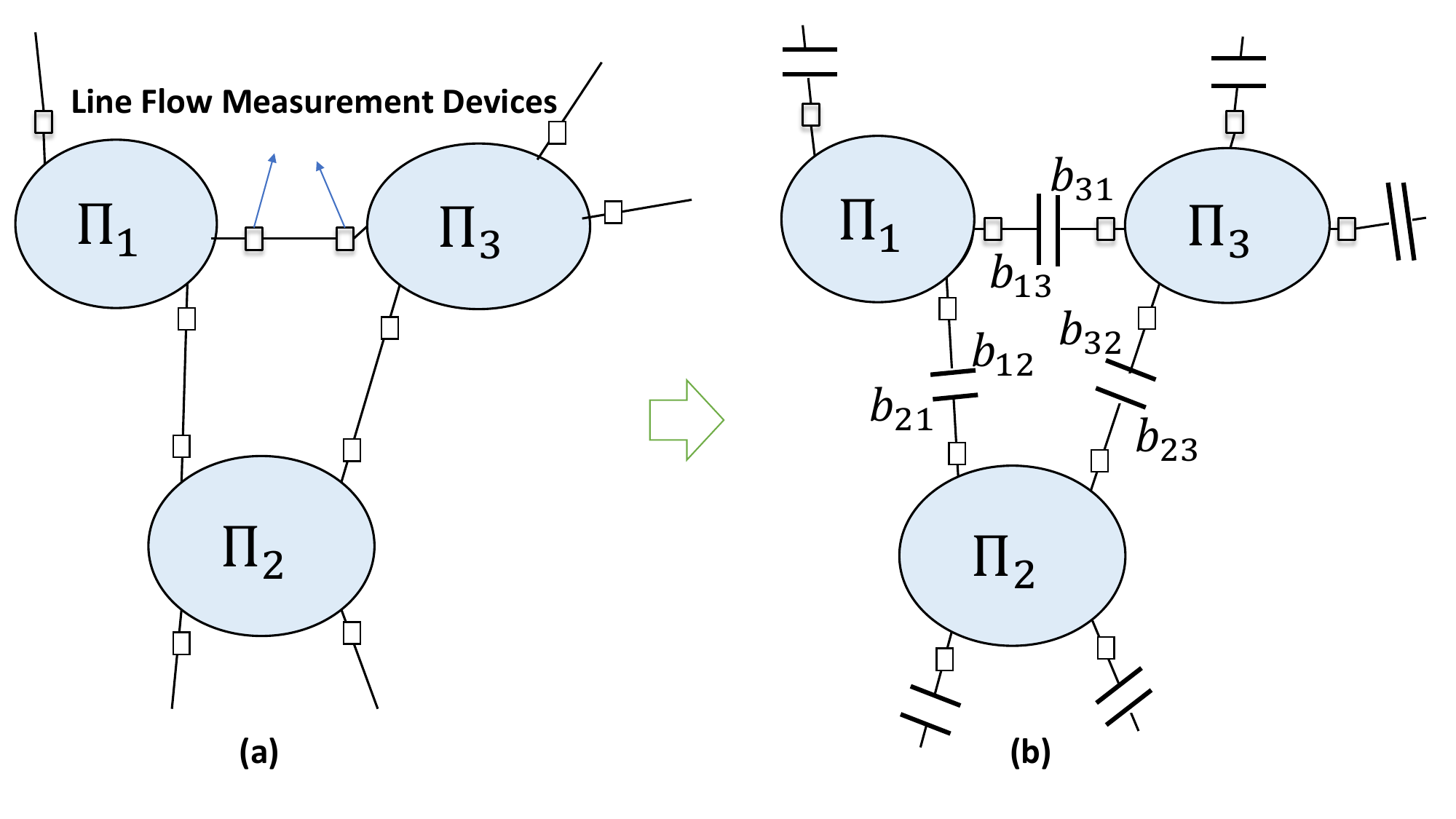}
\caption{(a) Schematic of  {modern power systems} with multiple prosumers. (b) Separation of neighboring prosumers.}
      \label{Separated}
\end{figure}

In \cite{engelmann2018toward}, the centralized OPF problem has been reformulated to a separable form suitable for distributed decomposition algorithms. For this purpose, the power system is divided into distinct local bus sets. As our objective is to monitor prosumers, each prosumer may be regarded as a local node that can be distinct from the neighboring nodes. Let $\Pi=\{ \Pi_1, \Pi_2, \dots, \Pi_n \}$ be the set of all prosumers and $\mathcal{N}_i = \{ j | \Pi_j \in \Pi , y_{ij} \neq 0 \}$ be the set of neighbors of prosumer $\Pi_i$, where $y_{ij}$ is the admittance of tie-line between prosumers  $\Pi_i$ and  $\Pi_j$. 
Following the procedure in \cite{engelmann2018toward}, auxiliary buses are introduced for tie-lines connecting adjacent prosumers. For example, for any $\Pi_j$ $\in \mathcal{N}i$, two buses, $b{ij}$ and $b_{ji}$, are added, with $b_{ij}$ connected to $\Pi_i$ and $b_{ji}$ connected to $\Pi_j$, as shown in Fig \ref{Separated}. The admittance between prosumers and auxiliary buses is $y_{ib_{ij}}=y_{jb_{ji}}=2y_{ij}$, and decoupled buses must have consensus values as formulated in (\ref{DOP5}).

As a result, the OPF problem of the reformed system can be written as in (\ref{DOPF})

\begin{subequations} \label{DOPF}
     \begin{align} \label{DOP1}
        {\min_{P, Q, \delta,V}} \quad & \sum_{\Pi_i \in \Pi} C_i(P_i)  
        \\ \notag
        {\rm{s.t}} \quad &
        V_i \sum_{j \in \mathcal{N}_i} V_{b_{ij}}(G_{i{b_{ij}}}cos(\delta_i-\delta_{b_{ij}})+B_{i{b_{ij}}}sin(\delta_i-\delta_{b_{ij}}))
        \\ \notag
        \quad &= P_i \;\; ; \; \forall \; \Pi_i \in \Pi
        \\ \notag \quad &
        V_i \sum_{j \in \mathcal{N}_i} V_{b_{ij}}(G_{i{b_{ij}}}sin(\delta_i-\delta_{b_{ij}})-B_{i{b_{ij}}}cos(\delta_i-\delta_{b_{ij}}))
        \\ \label{DOP2}
        \quad &= Q_i \;\; ; \; \forall \; \Pi_i \in \Pi
        \\  \quad & \label{DOP3}
        x_i^{min} \leq x_{i} \leq x_{i}^{max}; \;\; x=P,Q,V \;\; ; \; \forall \; \Pi_i \in \Pi
        \\  \quad & \label{DOP4}
        V_{s}=1 , \delta_{s}=0
        \\  \quad & \label{DOP5}
        \begin{bmatrix}
            \delta_{b_{ij}} \\ V_{b_{ij}} \\ P_{b_{ij}} \\ Q_{b_{ij}}
        \end{bmatrix}
        = 
        \begin{bmatrix}
             {\delta_{b_{ji}} }\\ V_{b_{ji}} \\ -P_{b_{ji}} \\ -Q_{b_{ji}} 
        \end{bmatrix} ; \;\; \; \forall \; \Pi_i \in \Pi \;and \; j \in \mathcal{N}_i
     \end{align}
\end{subequations}
where the local objective function of prosumer  $\Pi_i$ is denoted as $C_i$, with $V_i$, $\delta_i$, $P_i$, and $Q_i$ representing the equivalent voltage magnitude, voltage angle, net active power, and net reactive power of $\Pi_i$, respectively. Additionally, $G_{i{b_{ij}}}$ and $B_{i{b_{ij}}}$ signify the conductance and susceptance of the tie-line between prosumer $\Pi_i$ and auxiliary bus $b_{ij}$.
The constraints are outlined as follows:   {(\ref{DOP2})} corresponds to the power flow equations,  {(\ref{DOP3})} encompasses the operational bounds of the prosumers,  {(\ref{DOP4})} presents to the slack bus conditions, and  {(\ref{DOP5})} represents consensus constraints for neighboring prosumers \cite{engelmann2018toward}.  {Furthermore, while additional constraints such as transmission security can be incorporated into the DOPF formulation, maintaining consistency across all prosumers is essential.}  {Anomaly detection OPF (ADOPF) calculations are proposed to be conducted by neighboring prosumers} to facilitate the decentralized neighborhood watch-based monitoring  {of  {energy mismatch}.}

Consequently, it is necessary that each prosumer independently solves the ADDOPF problem for its neighbors. 
\cite{engelmann2018toward} describes a five-step algorithm based on augmented Lagrangian alternating direction inexact Newton (ALADIN) to solve (\ref{DOPF}) in a distributed fashion. 

At each iteration, steps should be performed locally and in parallel by prosumers as outlined in the following.
\begin{enumerate}
    \item Decouple (\ref{DOPF}) into local problems.
    Define $A_{ij}:=|\delta_{b_{ij}}-\delta_{b_{ji}}|+|V_{b_{ij}}-V_{b_{ji}}|+|P_{b_{ij}}+P_{b_{ji}}|+|Q_{b_{ij}}+Q_{b_{ji}}|$ as a penalty factor. Solve (\ref{LocalOPF}) below for all $\left\{\Pi_l | l \in \{ i \} \cup \{j | j \in \mathcal{N}_i \} \right\}$ in order to provide necessary information for both normal operation and monitoring the neighbors.
    \begin{align} \label{LocalOPF}
        {\min_{P_l, Q_l, \delta_l,V_l}} \quad & C_l(P_l) + \sum_{k \in \mathcal{N}_l} \lambda_k A_{lk}
        \\ \notag
        {\rm{s.t}} \quad & (\ref{DOP2}), \; (\ref{DOP3}), \; (\ref{DOP4})
    \end{align}
    \item Check if the solutions are consensus. 
    If  $|A_{ij}|<\epsilon$, the solution of step 1 is the final solution, and stop iterating, Otherwise, continue all of the steps below.
    \item Calculate the gradient and Hessian of the objective, along with the Jacobian of the constraint (\ref{DOP2}).
    \item Solve Consensus step problem as in \cite{wright2006numerical}.
    \item Line search for the globalization of the solution as in \cite{houska2016augmented}.
\end{enumerate}
Iterations persist until the solutions converge.

The solutions derived from the DOPF algorithm are regarded as reference power values that prosumers agree upon for production or consumption.  {Note that \cite{9363335} and \cite{9408219} offer potential avenues for integrating uncertainties in communication systems and power variations in renewable energy resources, respectively. However, since our goal is to develop a monitoring scheme, their integration falls outside the scope of this paper.}

The true power value of each prosumer can be obtained by utilizing measurement, using hardware accessible to individual prosumers, namely power flow measurement.

In general, $\Pi_i$ may have multiple connections (tie-lines) to other prosumers,  as illustrated in Fig. \ref{Separated}(a), where direct measurement is not applicable. However, most of the prosumers in MPS are equipped with power flow measurement devices.

 {Let  $p^{ref}_{ij}(z)$ be the reference power of 
prosumer  $\Pi_i$ that is calculated by neighboring prosumer $\Pi_j$ at the time interval $z$ within each $\tau$-minute time interval such that , $t \in [z \tau, (z+1) \tau)$. Furthermore, let the rate of measurement and data transmission be $\mathcal{L}$ samples in $\tau$ minutes. Thus, the difference between the energies of agreed-upon power and the actual power of $\Pi_i$ can be calculated by neighbors as }
\begin{equation*} \label{Eq_smartmeter}
d_{ij}(z) = \left( \sum_{k \in \mathcal{N}_i}  \sum_{l =1 } ^{\mathcal{L}} p^{probe}_{ik}(l) \right) - \tau p^{ref}_{ij}(z) \; ,  \;\;\; \forall j \in  \mathcal{N}_i.
\end{equation*}

Each prosumer can monitor the behavior of neighbors by calculating  {$d_{ij}(z)$} at each time interval. The prosumers can acquire measurement data from the neighbors of their neighbors through a two-hop communication process. 
 {By proper choice of $\tau$, frequent variations of resources and loads could be captured by reference power and probe power values and distinguished if it is an anomaly or not.}
Only those prosumers that continuously deviate from their reference power values are considered \emph{anomalous} agents.
Note that, in this study, our focus is only on anomalies where the malicious sources do not follow the protocol. 

\section{Automated Detection and Resolution of anomaly} \label{Section3}
Recalling from the previous section, we can derive reference power for all neighbors of a prosumer via ADOPF calculations. Also, we can calculate the  {energy mismatch} of each prosumer from its reference power by the distributed dynamic probing technique. to distinguish the disturbance of power generated or consumed by each prosumer caused by system changes from the malicious anomalies, a dead zone filter is applied to $d_{ij}(t)$ values, preventing false positives.
A significant and persistent deviation indicates an anomaly, prompting necessary action to mitigate the prosumer's impact on the network.

In MPS, both the act of anomaly and the magnitude and duration of the anomaly are critical. 
Thus, it is necessary to make fast reactions for severe anomalies, such as isolating the anomalous prosumer or disconnecting the anomalous prosumer from the grid.
For milder instances of anomaly, such as accidental power deviations or minor deviations from the scheduled values, penalties as a measure can be pursued to address and mitigate the behavior of the non-compliant prosumer.

 {Another important consideration is the severity of the anomaly. It involves determining whether the mismatch is increasing or decreasing in relation to the system's responses.}
Enforcing penalties for prosumers displaying an uptick in anomaly is crucial, while concurrently offering incentives to those prosumers who actively mitigate their anomaly is equally important.


 
To fulfill the aforementioned objectives, the anomaly factor, ${F_{ij}}$, is defined as follows: 
\begin{equation}
    {F_{ij}(k)} = \frac{1}{N(d(k))}\left( F_{ij}(k-1) + d_{ij}(k ) D_{ij}(k) \right)
\end{equation}
where $N(d(k))$ is the factor for returning to normal operation and adjusting the speed of reacting to anomaly. 

\begin{equation}
    N(d(k))= 
    \begin{cases}
        1 , & \text{if}\ |d(k)|> \epsilon \\
        N_0>1,  & \text{Otherwise}\
    \end{cases}
\end{equation}
 {where determining the appropriate value of $\epsilon$ is crucial to determine whether deviations beyond this threshold pose risks to the stability or efficiency of the system. Therefore, $\epsilon$ should be carefully determined during the system planning phase, and prosumers should be aware that exceeding this threshold may result in penalties.}
Upon cessation of anomaly by the prosumer, where $\frac{1}{N}<1$, the values of $F_{ij}$ begin a gradual decline over time, leading the system to revert to its normal operational mode.
Additionally, $D(k)$ is defined as
\begin{subnumcases}{D(k)=}
    a(d(k)- d(k-1)) & \text{if}\ $d(k) \neq d(k-1)$ \label{Rate1}
    \\
    1, & \text{if}\ $d(k) = d(k-1) > 0$ \label{Rate2}
    \\
    -1, & \text{if}\ $d(k) = d(k-1) < 0$ \label{Rate3}
\end{subnumcases}
where (\ref{Rate1}) implies the rate of anomaly where the prosumer is adjusting the power generation.
%
The control variable $a>0$ is employed to modulate the impact of the rate of anomaly changes in the computation of the anomaly factor.
Equations (\ref{Rate2}) and (\ref{Rate3}) imply that if the prosumer continues anomalous at a constant value, $F_{ij}(k)$ increases with respect to the value of $d_{ij}(k)$, thus, $D_{ij}(k)$ gets unit value matching the sign of $d(k)$.

To mitigate accidental anomalies, a penalty is assigned to the anomalous prosumers. 
The penalty, $P_i(t)$, for prosumer $\Pi_i$ during time interval $t \in [(k-1) \tau,k \tau)$, is determined by aggregating penalties from its neighbors. The average of these penalties serves as the penalty price for prosumer $\Pi_i$.

A penalty price function is proposed, modeled by an exponential growth function of the anomaly factor. This design ensures that the penalty steadily increases in proportion to the persistence of anomalies within the system. Let
\begin{equation} \label{PenaltyOperator}
    P_{i}(t)=\frac{1}{|\mathcal{N}_i|} \sum_{j \in \mathcal{N}_i} P_{ij}(t)
\end{equation}
and 
\begin{equation} \label{PenaltyProsumer}
     P_{ij}(t)= (c^{F_{j,i}(k)}-1)
\end{equation}
where $c>1$ is the penalty factor.
Moreover, the growth function is subtracted by 1 to keep the penalty price value at zero during the normal operation condition. Neighboring prosumers report the calculated penalty values by (\ref{PenaltyProsumer}) to the operator and the penalty of the anomalous prosumer is determined by (\ref{PenaltyOperator}).

The penalty value gradually diminishes over time as the prosumer ceases anomalous behavior, a reduction that may stem from prior penalization at a specific point in time.
Conversely, in the event of continued anomaly such as malicious anomaly, the penalty value continues to escalate.
To cope with this case, the penalty value calculated at each neighbor via (\ref{PenaltyProsumer}), is compared to a threshold margin, $C_{th}$, and the system continues normal operation in time interval $t \in [k \tau, (k+1) \tau)$ while $P_{ij}(t) \leq C_{th}$. Nevertheless, when $P_{ij}(t) > C_{th}$,  $\Pi_j$ detects that $\Pi_i$ is  {malicious anomaly and reports it to the utility. The mitigation action considered in this study is to isolate the particular prosumer similar to \cite{EMS1} until the malicious anomaly can be resolved.}
 {Note that the choice of the threshold value ($C_{th}$) affects the likelihood of miss detection (failure to detect) and false alarms (unwanted detection). Specifically, by increasing $C_{th}$, the possibility of miss detection would increase, and by decreasing it the possibility of false alarm would increase. Meanwhile, the algorithm is designed to consider the duration of the anomaly as well. For instance, if the $C_{th}$ value is high, the prosumers would continue anomalous behavior and the detection would eventually take place over time. To sum up, the choice of the $C_{th}$ depends on the conditions of the system and it requires periodic re-evaluation based on system conditions, historical data, and stability analysis.} 

\begin{figure}[t]
    \includegraphics[width=1\linewidth ,trim={0 2cm 0 2cm},clip]{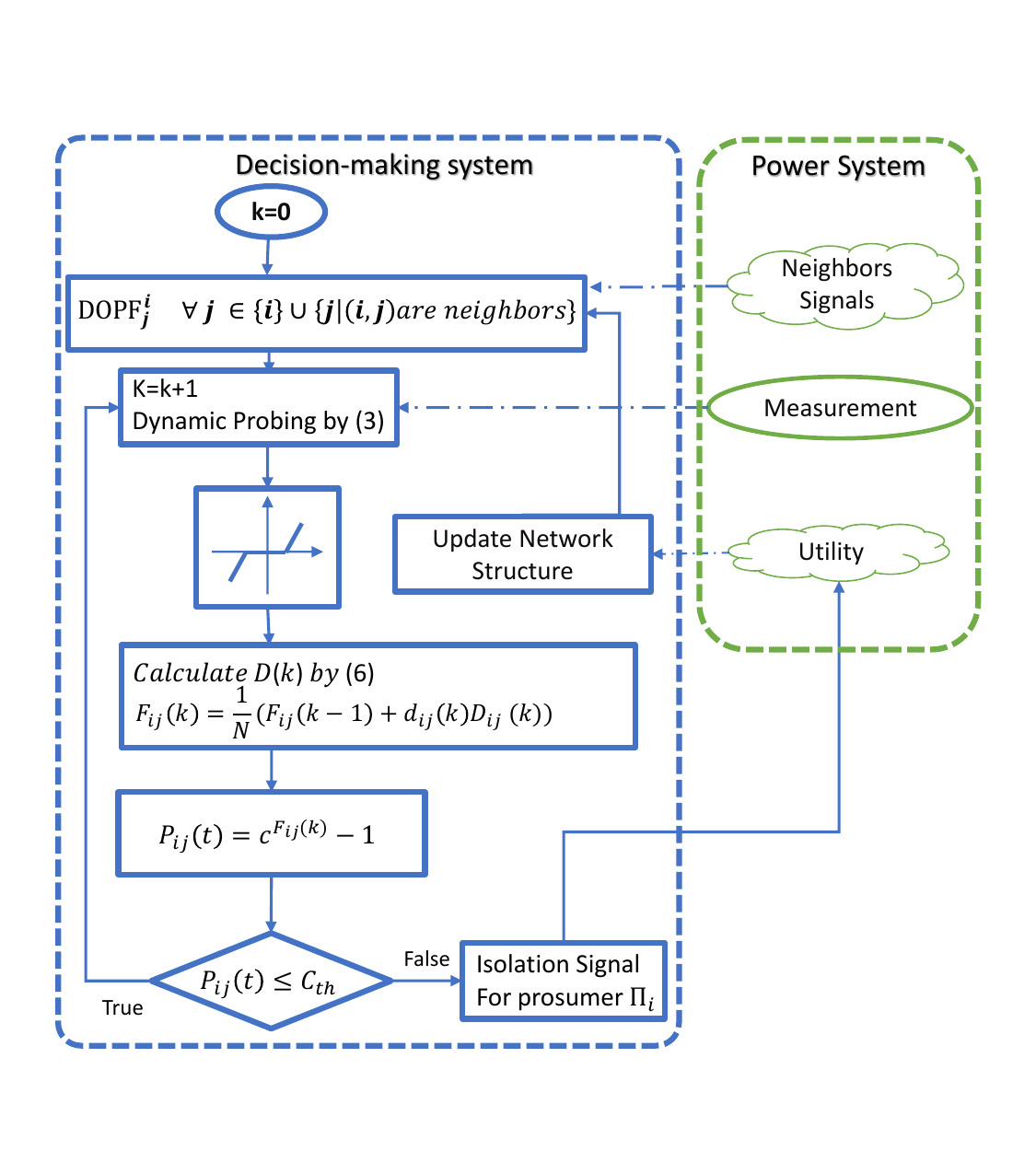}
    \centering
    \caption{ {Flowchart of anomaly detection and mitigation algorithm.}}
    \label{FlowChart}
\end{figure}

 {While monitoring operates in a distributed manner, a higher-level enforcement agent such as utility oversees the isolation of anomalous prosumers and network topology changes. The utility's presence complements distributed monitoring by serving as a privileged entity, ensuring effective coordination without disrupting the distributed nature of the monitoring process.} If more than a certain ratio of the neighbors of a certain prosumer sends an isolation signal, the utility can isolate the prosumer from the network until the source of the anomaly is removed. During the detection and isolation of a prosumer, the utility can modify the network structure, requiring updates to neighbor sets and factor values.
The proposed algorithm is summarized in Figure \ref{FlowChart} for more clarification. 

\section{Simulation}\label{Section4}

\begin{figure}[t]
    \centering
    \includegraphics[width=0.9\linewidth, trim={0 0 0.1cm 0},clip ]{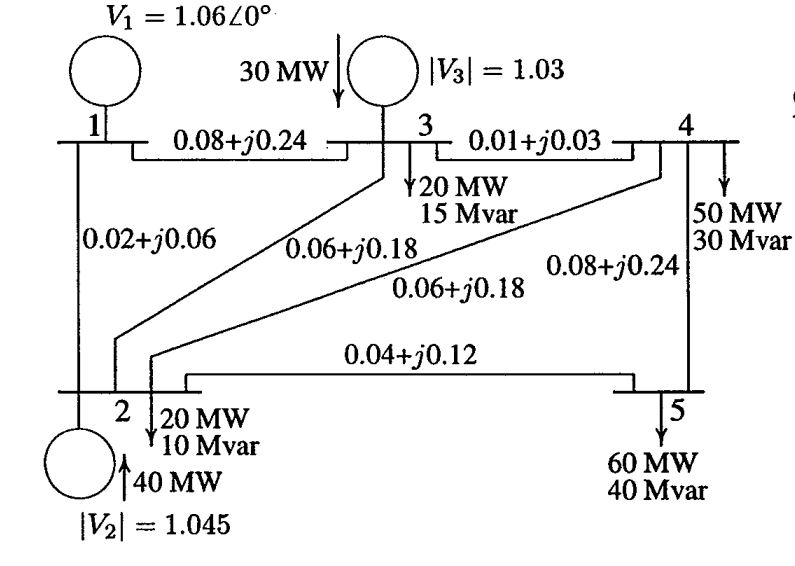}
    \caption{Schematic of the modified IEEE 5-Bus System and initial power flow values \cite{saadat1999power}.} \label{5bus} 
\end{figure}

To assess the efficacy of the suggested distributed detection and mitigation algorithms, we utilize a modified version of the IEEE 5-Bus system (see Fig. \ref{5bus}), as introduced in \cite{saadat1999power}.

 {Each substation has been considered as a prosumer.}
Since all prosumers are within the two-hop neighborhood of each other, they can calculate reference power for all prosumers in the networks. 
 {According to the information provided in \cite{saadat1999power} such as generation cost of generators, results of DOPF problem would be calculated $p^{ref}_1=p^{ref}_{(j,1)}=23.56 MW$, $p^{ref}_2=p^{ref}_{(j,2)}=49.56 MW$, and $p^{ref}_3=p^{ref}_{(j,3)}=39.04 MW$ for $j=1,2,\dots,5$. as the net power of each node. Since our focus is on the anomaly detection problem, we assume that the DOPF results are given and remain fixed. Also, \cite{saadat1999power} considers $\Pi_1$ as a slack bus, which compensates for power balance in case of any isolation of malicious anomaly.}

Consider the system to be operating for $t \in [k \tau, (k+1) \tau$ for $k=1,2,\dots,24$ with fixed load values.  {Since the power values are constant, $\tau$ is considered intervals with a few minutes duration. In this problem, we have considered 2 hours of network operation with $\tau=5$ minutes. Also, energy value and power values would be linearly correlated and could be replaced with each other.} The control parameters of the algorithm are chosen to be $N_0=3$, $C_{th}=1300$, $c=1.06$, $a=1$, and $\epsilon=0.1MW$. 

{ {
Now assume that $\Pi_2$ is subjected to two distinct instances of anomalies, accidental and malicious; resulting in energy mismatch.}}
First anomaly occurs at $k=4$ where power generation of the prosumer increases $15\%$. This causes the anomalous factor and penalty value to increase for $\Pi_2$ at $k=4$. Therefore, the anomaly is removed at $k=5$. 
Therefore, the anomalous factor and penalty price diminish proportionately, facilitating the prosumer's return to a normal operational condition.

Next, another anomaly occurs at $k=8$ which causes the prosumer to produce $10\%$ less than reference power. Since anomaly is less severe in this scenario compared to the first anomaly, the anomalous factor and penalty values start to have lower values. 

However, in contrast to the preceding scenario, the anomaly persists until the penalty value reaches the threshold, leading to the decision to isolate $\Pi_2$ at $k=22$  {which leads to $p^{ref}_1=83.01MW$, $p^{ref}_2=-20MW$, and $p^{ref}_3=50MW$.}

\begin{figure}[t]
    \centering
    \includegraphics[width=1\linewidth ,trim={0 1cm 0 1cm},clip ]{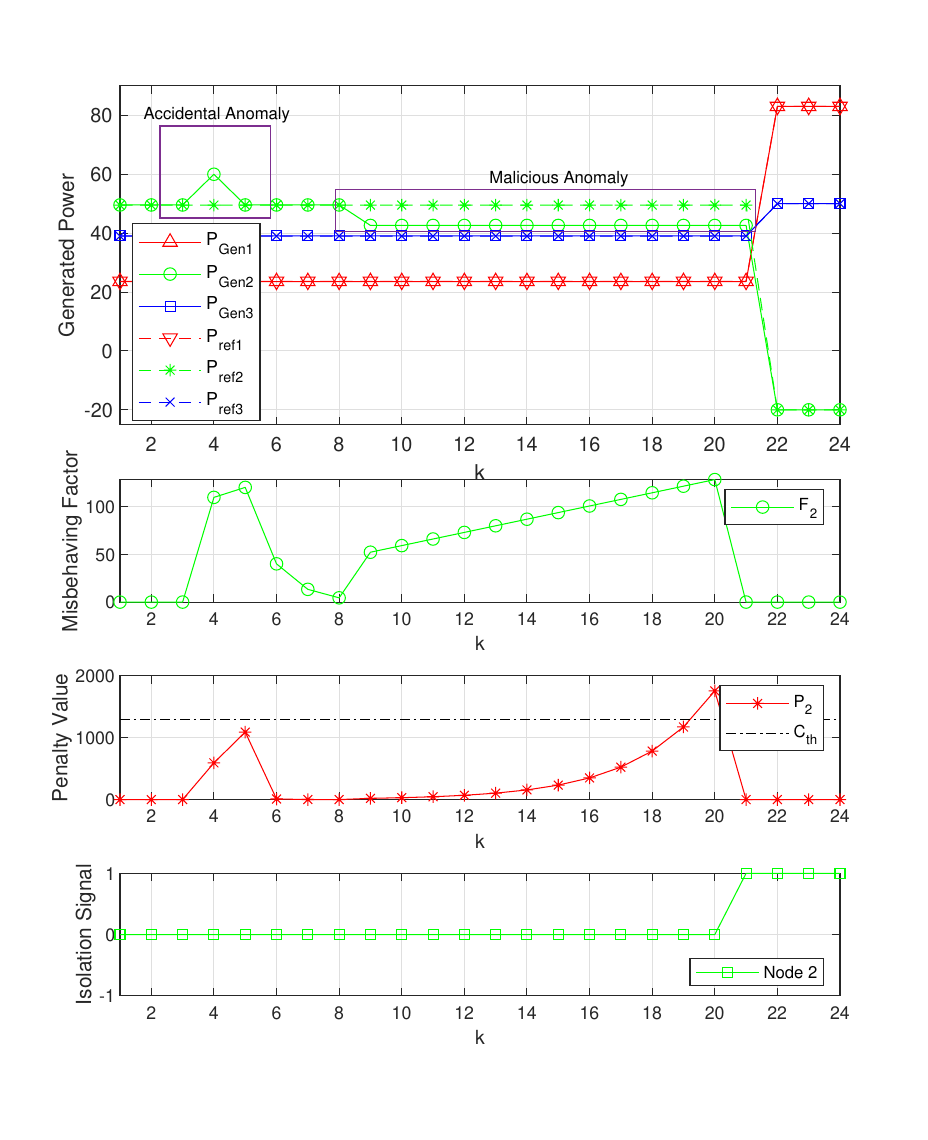}
    \caption{ (a) Generated power, (b) anomalous factor values of node 2 for $k \in \{1,2,\dots,24\}$, (c)Penalty value of node 2, and (d) isolation signal.}  \label{results} 
\end{figure}

Figure \ref{results} depicts the variations in power generation, anomalous factor, and penalty value for both scenarios.


For the sake of simplicity, the values are depicted based on a single neighbor's calculation. It can be assumed that all neighbors will independently reach the same calculations locally.
 
\section{Conclusion}\label{Section5}
This paper presents novel distributed dynamic probing-based algorithms for detecting and mitigating anomaly in prosumer-based MPS. The probing technique utilized tie-line power flow measurements as a foundation for effective anomaly detection.
An automated detection and resolution algorithm is formulated, where a anomaly factor is introduced to quantify the severity, duration, and rate of  {energy mismatch} caused by anomalys. The anomaly factor is calculated by neighbors of the anomalous prosumers forming a neighbor watch-based detection algorithm.


Next, a mitigation approach is proposed by imposing penalties on anomalous prosumers, isolating malicious anomalous prosumers from the network, and updating the network operation with a new topology. In order to calculate penalties for the anomalous prosumers, an exponential growth function based on the anomalous factor is proposed. Simulation results demonstrate the effectiveness of the proposed system and algorithm.

Broadening the scope of the proposed approach entails addressing challenges such as the impact of transmitted parameter noise and the uncertainties associated with the dynamic probing algorithm. Investigating the convergence of the algorithm under such uncertainties is a crucial avenue for future research.

\bibliographystyle{ieeetr}
\bibliography{references}

\end{document}